**Periodic chiral magnetic domains in single-crystal nickel nanowires**

Jimmy J. Kan[1,*], Marko V. Lubarda[1,3], Keith T. Chan[1,†], Vojtech Uhlir[1], Andreas Scholl[2], Vitaliy Lomakin[1], and Eric E. Fullerton[1]

1. Center for Memory and Recording Research, University of California – San Diego, La Jolla, California 92093, USA
2. Advanced Light Source, Lawrence Berkeley National Laboratory (LBNL), 1 Cyclotron Road, Berkeley, California 94720, USA
3. Faculty of Polytechnics, University of Donja Gorica, Oktoih 1, 81000 Podgorica, Montenegro

ABSTRACT

We report on experimental and computational investigations of the domain structure of ~0.2 x 0.2 x 8 μm single-crystal Ni nanowires (NWs). The Ni NWs were grown by a thermal chemical vapor deposition technique that results in highly-oriented single-crystal structures on amorphous SiOx coated Si substrates. Magnetoresistance measurements of the Ni NWs suggest the average magnetization points largely off the NW long axis at zero field. X-ray photoemission electron microscopy images show a well-defined periodic magnetization pattern along the surface of the nanowires with a period of $\lambda = 250$ nm. Finite element micromagnetic simulations reveal that an oscillatory magnetization configuration with a period closely matching experimental observation ($\lambda = 240$ nm) is obtainable at remanence. This magnetization configuration involves a periodic array of alternating chirality vortex domains distributed along the length of the NW. Vortex formation is attributable to the cubic anisotropy of the single crystal Ni NW system and its reduced structural dimensions. The periodic alternating chirality vortex state is a topologically protected metastable state, analogous to an array of 360° domain walls in a thin strip. Simulations show that other remanent states are also possible, depending on the field history. Effects of material properties and strain on the vortex pattern are investigated. It is shown that at reduced cubic anisotropy vortices are no longer stable, while negative uniaxial anisotropy and magnetoelastic effects in the presence of compressive biaxial strain contribute to vortex formation.

The investigation of magnetism in mesoscale structures has attracted considerable interest in recent years[1-6]. As the structural dimensions of materials are reduced down to typical length scales associated with ferromagnetic ordering, competitions between several magnetic interactions arise and can result in the formation of new, intricate magnetic configurations[7-11]. These new structures could offer a pathway towards future applications in high-density data storage[12-16], compact magnetic sensors[17, 18], high frequency nanoscale oscillators[19, 20], and magnetic logic[21-24]. For example, a competition between the exchange interaction, magnetic anisotropy, stray field energy, and geometric confinement leads to the emergence of vortex structures in magnetic disks[25, 26] and stripe domains in thin films[27, 28]. Magnetic vortices are flux closure states that exist in micron and sub-micron diameter magnetically-soft disks. The vortex state is characterized by the circulation of the magnetization, either counterclockwise or clockwise ($c$=+1 or -1), and the polarity of the core (the magnetic singularity at the center of the disk), which points either up or down ($p$=+1 or -1) perpendicular to the disk plane. The combination of circulation and polarity defines the vortex chirality either right handed ($cp$=+1) or left handed ($cp$=-1). Stripe domains conversely arise in thin films with perpendicular magnetic anisotropy as originally described by Kittel[29-31]. The width of the meandering domains is determined from a balance between exchange energy (i.e. domain wall energy) and dipole energy and varies with film thickness. The morphology of the domains depends on the field history. In this paper, we describe the domain structure that arises in single-crystal nickel nanowires (NWs). In these NWs, the combination of the well-defined magneto-crystalline anisotropy and NW shape results in the emergence of a periodic vortex domain structure where the vortex chirality alternates handedness along the primary axis of the wire.

Previous investigations of magnetic NWs have typically been carried out by studies on amorphous or polycrystalline wires grown by nanoporous template electrodeposition or self-assembly[32-35]. These methods of synthesis induce crystalline grains in the microstructure that can cause local variations in the magnetic anisotropy and moment. Thus it is impractical to explore the full magneto-crystalline anisotropy (MCA) of such structures, and as a result, these previous studies have largely varying conclusions depending on the differences in crystalline quality[36-40]. For these types of cylindrical nanowires, a quasi-1D model where the magnetic response is dominated by the shape anisotropy is typically sufficient to explain the magnetic properties. Because of the high aspect ratio, the static and magnetic reversal properties can be modeled using a Stoner-Wohlfarth or curling-mode model assuming a single, strong shape anisotropy along the wire axis[41, 42]. This strong shape anisotropy leads to magnetic reversal that is characterized by domain nucleation and propagation from the wire ends.

The Ni NWs described in this study are synthesized by a thermal chemical vapor deposition (CVD) technique that results in highly-oriented, single-crystal Ni NWs grown on amorphous SiOx coated Si substrates. Details of this growth technique have been published previously[43]. The resulting NWs have square cross sections with about 200 nm width, and lengths of up to 8 μm. Transmission electron microscopy images and electron diffraction patterns in Fig. 1 show the long/growth axis of the NWs are along the <001> direction and the side of the NWs are bounding surfaces that are atomically smooth. Analysis by coherent x-ray diffraction has shown the NWs to be single-crystalline[44,45].

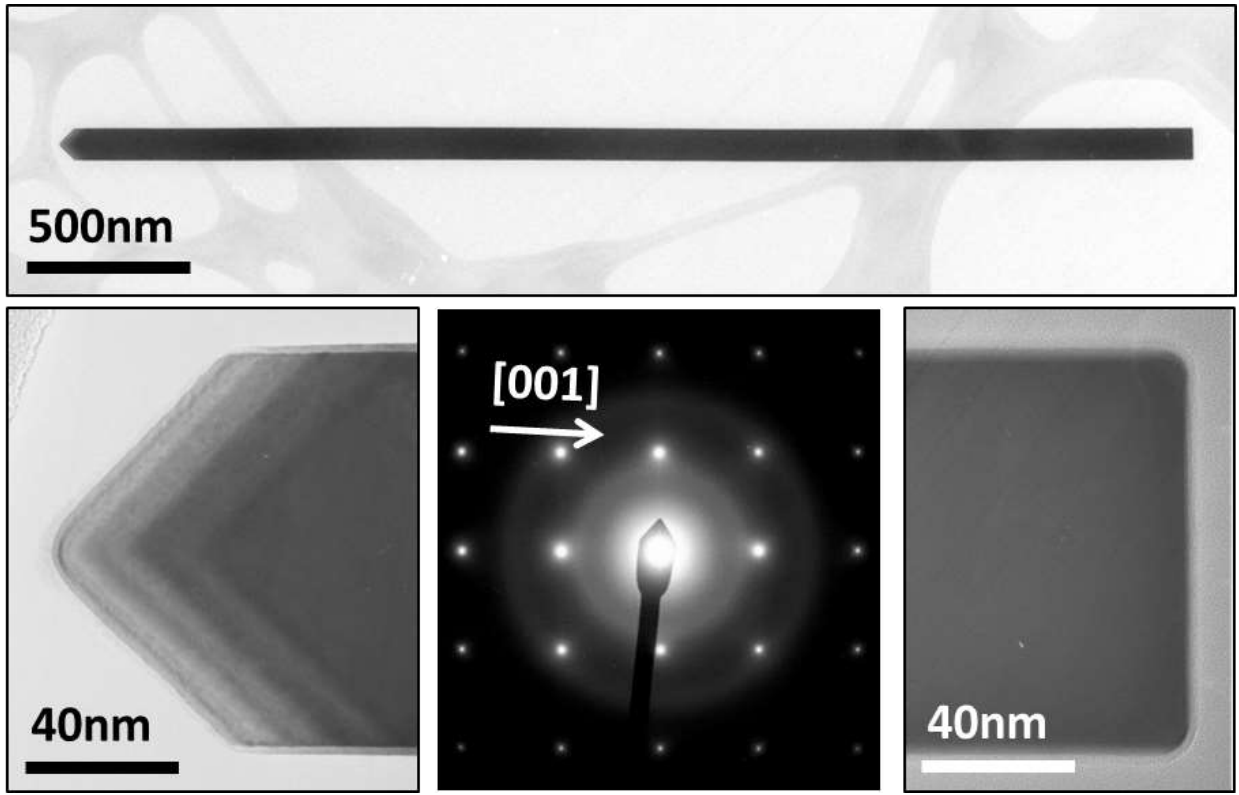

*Figure 1: Transmission electron microscopy images and electron diffraction patterns of a Ni NW after transfer to TEM grid.*

For nickel single crystals, the MCA easy axis is the <111> direction[46]. At any point inside the crystal, there are eight equivalent <111> easy directions at angles of approximately 55° relative to the wire axis as shown in Fig. 2a. The MCA is in competition with the magnetic shape anisotropy, which prefers the magnetization to lie along the nanowire's long axis (<001> directions). Previous measurements of the anisotropic magnetoresistance (AMR) response of the Ni NWs at $T$ = 10 K reflect this competition of magnetic energies[43]. In AMR measurements, the electrical resistance of a ferromagnetic structure depends on the relative angle between the direction of current flow and average magnetization direction. The value of the resistivity is given by $\rho(\theta) = \rho_\parallel - (\rho_\parallel - \rho_\perp)\sin^2(\theta)$ where $\rho_\parallel$ and $\rho_\perp$ represent the resistivities for configurations fully magnetized parallel and perpendicular to the wire axis, respectively. As shown in Fig. 2b, the resistance at remanence is intermediate between $R_\parallel$ and $R_\perp$, suggesting that the average magnetization at $H$ = 0 lies off-axis. The AMR curves of the Ni NWs contain only small regions of hysteresis, implying that this intermediate magnetization state is reachable regardless of initial magnetic conditions. The shape of these field angle and magnitude dependent AMR curves suggest a competition between the shape anisotropy and MCA.

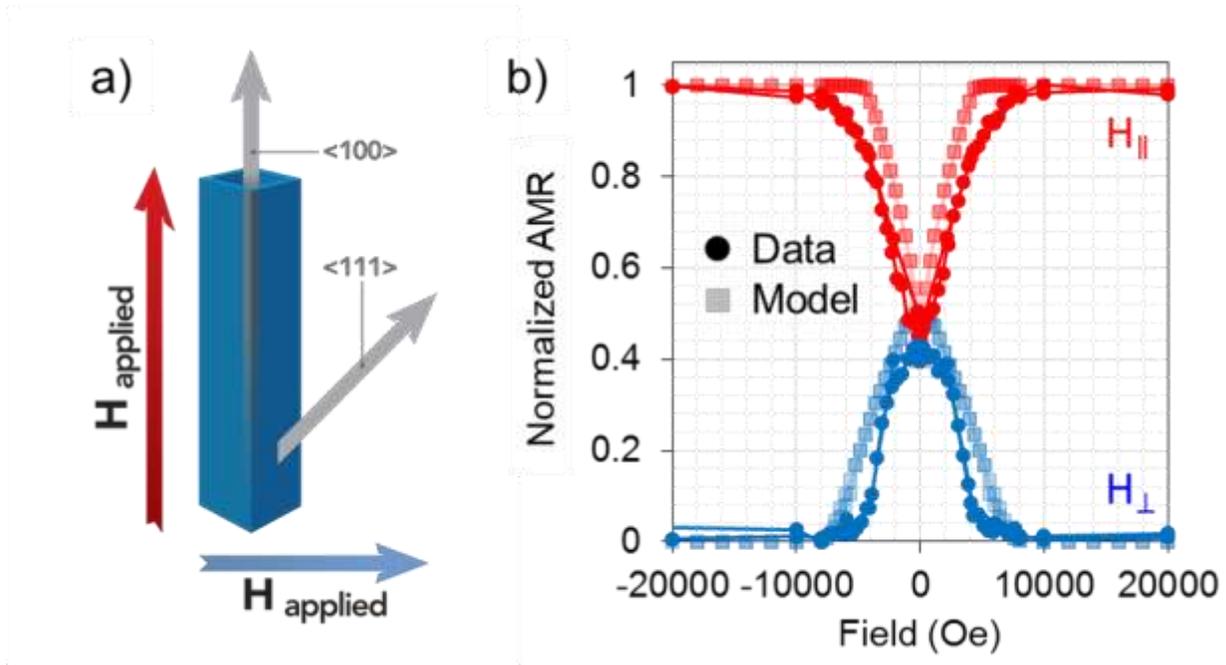

*Figure 2: a) Diagram of Ni NW structure with example crystalline axes indicated. 8 equivalent <111> axes exist. b) Anisotropic magnetoresistance behavior of Ni NW with varying external applied field orientations, and AMR calculated from micromagnetic modeling.*

The limitation of AMR measurements is that they only measure the average overall magnetization of the NWs. In order to visualize the space resolved micromagnetic configurations, we have performed magnetic and structural imaging using x-ray photoemission electron microscopy (PEEM). For PEEM imaging, the NWs were transferred onto silicon oxide substrates by a contact transfer technique. To prevent substrate charging during imaging, substrates were coated with 10-nm Pd prior to NW transfer. Imaging was performed at room temperature on EPU beamline 11.0.1 at the Advanced Light Source (Lawrence Berkeley National Lab) using PEEM3. PEEM3 has an energy range of 150-2000 eV with a spatial resolution down to 30 nm, and offers full polarization control. Magnetic contrast in photoemission experiments is provided by an asymmetry in the absorption cross sections for left and right circularly polarized x-rays for ferromagnetic materials. This dichroism allows for the direct imaging of the local magnetization vector when coupled to a spatially sensitive detector.

A schematic of the imaging experiment is shown in Fig. 3a. Incident x-rays are turned to the Ni $L_3$ and $L_2$ absorption edges at 853 eV and 871 eV, respectively, and arrive at the sample surface with an angle of approximately 30° relative to the substrate surface. To observe the off-axis magnetization components suggested by AMR measurements, the NW is oriented at an almost orthogonal angle relative to the incoming x-ray direction (~82°). This almost orthogonal configuration maximizes the magnetic contrast as XMCD is sensitive to the magnetization component along the x-ray propagation axis. X-ray absorption spectra (XAS) and XMCD from the wire are shown in Fig. 3b. The $L_2$ absorption edge exhibits a slight multiplet splitting, indicating a possible surface oxidation of the NW. Besides this slight surface oxidation, the XAS behavior is typical of pure Ni without impurities.

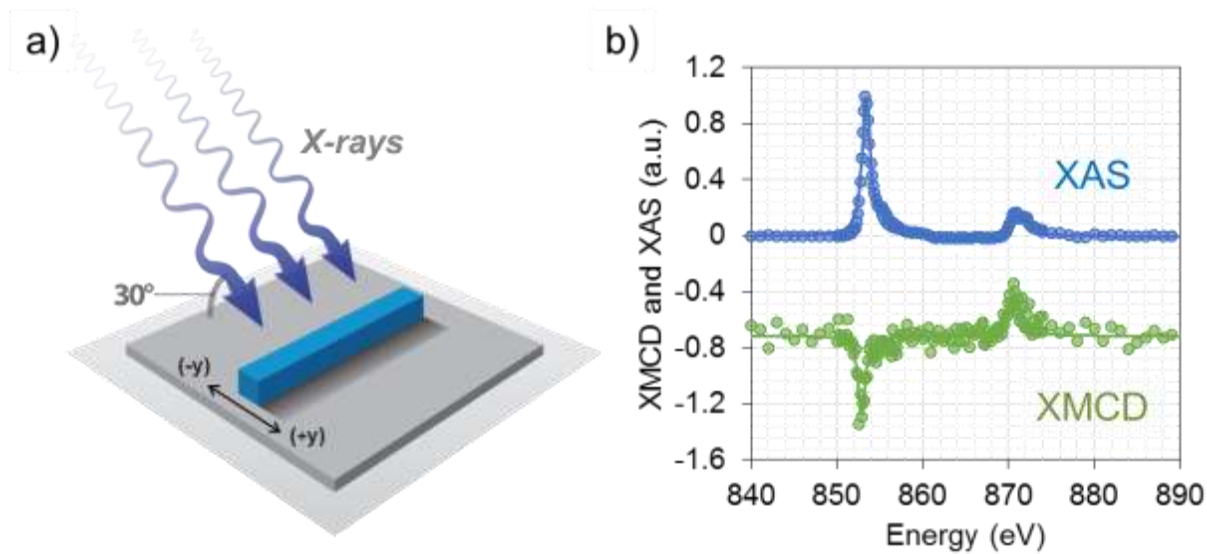

Figure 3: Diagram of NiNW structure with example crystalline axes indicated. 8 equivalent <111> axes are oriented ~53° relative to the <001> axes.

X-ray magnetic circular dichroism (XMCD) is defined as the ratio of $XAS(RCP)/XAS(LCP)$ where RCP and LCP are right circular and left circular polarized light respectively. To observe the magnetic configuration of the Ni NW, XMCD images are recorded at the Ni $L_3$ edge (853 eV). XAS and XMCD images from a Ni NW are shown in Fig. 4. The brightly lit section of the image in Fig. 4a corresponds to the structure of the wire, whereas the dark section is a shadow arising from the obscured x-ray flux due to the physical size/height of the NW. The probing depth of the PEEM technique is limited by the inelastic mean free path of photoelectrons, which is 1.6 nm. As a consequence, XMCD is a surface sensitive technique that can only provide information about the magnetization of the top 5 nm of the sample surface. Though the electron inelastic mean free path is small, the x-ray attenuation length from absorption is one order of magnitude larger, approximately 40 nm at this absorption edge. Due to the small thickness of the NWs, some residual flux of x-rays will traverse the thickness of the wire and contribute to photoexcitation of electrons from the conductive substrate, forming a "shadow" region. The integrated absorption through the body of the nanostructure is also governed by XMCD effects and can thus give an indication of the bulk magnetization configuration by examining the XMCD of the shadow[47] (even though no magnetic material is physically present in that region). Figure 4b shows an image of the wire and shadow with XMCD contrast. The bright and dark regions indicate local magnetization components parallel (+y) and anti-parallel (-y) to the x-ray direction.

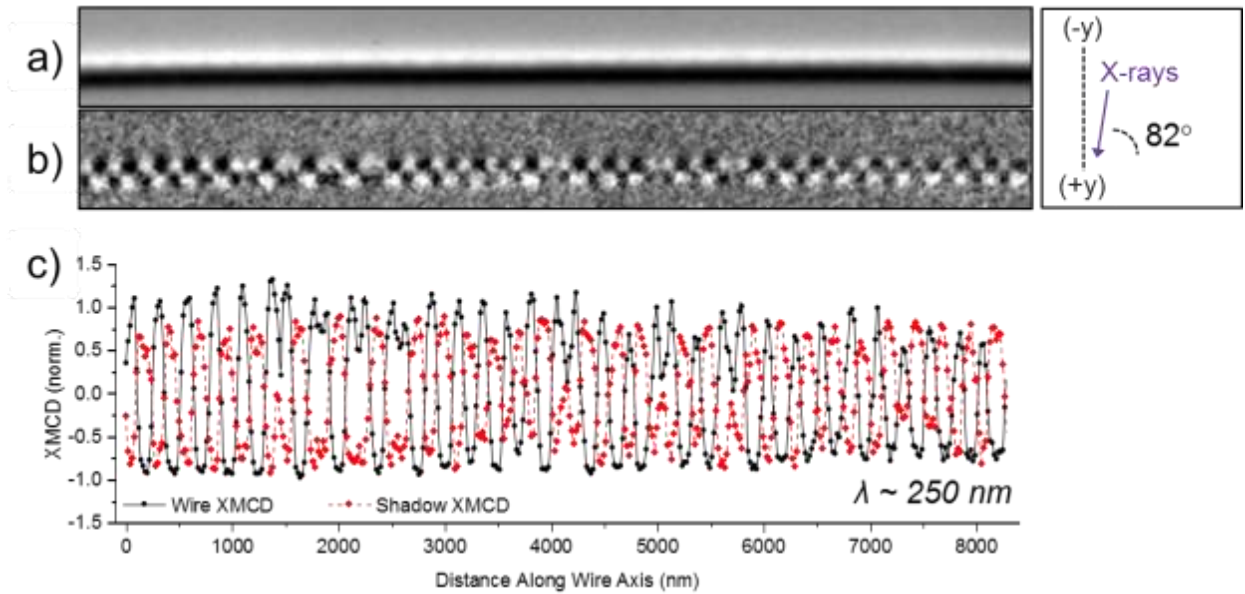

Figure 4: a) XAS image of a Ni NW. b) XMCD image of the Ni NW with magnetic contrast. c) Horizontal line scans of wire and shadow images with XMCD contrast. Inset) Description of axes and experimental geometry

The XMCD images show a clear periodic domain pattern down the length of the NW. A line intensity scan along the NW and shadow XMCD images in Fig. 4c shows well-formed oscillations in the magnetization with a period of $\lambda \sim 250$nm. Due to the XMCD absorption within the NW as described above, the also-periodic magnetization configuration of the shadow confirms that the domain pattern is not only existing on the surface of the NW, but persists in the NW bulk. The shadow profile is almost out-of-phase with the surface profile, suggesting that the bulk component roughly follows the surface magnetization. For example, if the local surface magnetization in one area of the wire is oriented collinear to the x-ray flux direction (+y), the surface XMCD image would be locally bright in this area due to strong absorption. If the local core magnetization is similarly oriented along (+y), then a strong absorption of the x-rays would reduce the transmitted flux to the substrate surface, resulting in a dark area. From these images, it is apparent that the bulk and surface magnetizations of the NW are mutually periodic along the length of the NW.

To better understand the images and the underlying magnetic configuration, finite element micromagnetic simulations were performed to computationally investigate the response of the single crystal Ni NW. The simulation results were obtained by numerically solving the Landau-Lifshitz-Gilbert (LLG) equation using the FastMag Micromagnetic Simulator[48] for a NW with the magnetic properties of bulk Ni fcc crystals. AMR loops, magnetization hysteresis loops, and domain images were computed for a 200 x 200 x 8000 nm NW discretized using tetrahedral elements of 10 nm nominal edge size. The values of material parameters of the NW used for the first set of simulations were: saturation magnetization $M_S$ = 480 emu/cm$^3$, exchange stiffness constant $A$=1.0 μerg/cm, first and second order cubic anisotropies $K_1$ = -1.20 Merg/cm$^3$, $K_2$ = 0.41 Merg/cm$^3$, respectively, which correspond to values expected at $T$ = 10 K[46]. Negative first order anisotropy ($K_1$ < 0) implies that the easy axis of magnetization is as along the <111> directions. Figure 2b shows the field and angular dependence of the AMR calculated based on micromagnetic modeling, which demonstrates good agreement with the experimental measurements. The key features of intermediate resistance remanent state, high saturation field, and low hysteresis are reproduced in simulations.

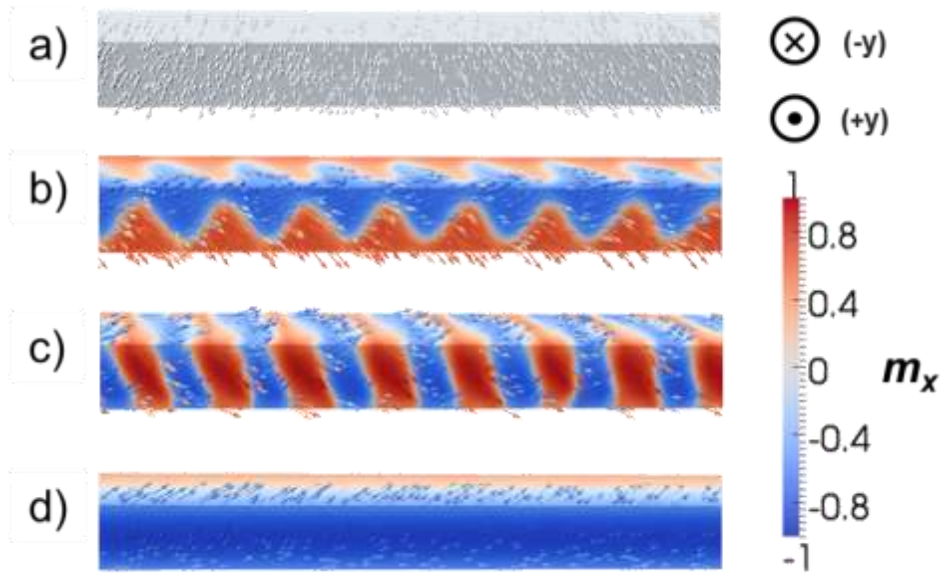

Figure 5. Magnetization configurations of Ni NW at $T$ = 10 K obtained by micromagnetic simulations. (a) Saturated state with magnetization uniformly pointing in x-direction in the presence of saturating magnetic field of 10 kOe. (b) Oscillatory precursor state with period of λ = 240 nm which develops when field is gradually reduced from saturating value (10 kOe) to 2.5 kOe. (c) Periodic alternating chirality vortex state with period of λ = 240 nm that evolves from precursor state if applied field is abruptly removed. (d) Uniform chirality vortex state obtained from precursor state following gradual reduction of applied field from 2.5 kOe to zero.

Figure 5 illustrates magnetization configurations of the cubic anisotropy Ni NW obtained by micromagnetic simulations under different applied magnetic field conditions at $T$ = 10 K. A uniform 10 kOe magnetic field applied perpendicular to the wire long axis uniformly saturates the NW magnetization in the (+$y$) direction as shown in Fig. 5a. As the field is reduced to the vicinity of 2.5 kOe, a precursor oscillatory state develops (Fig. 5b) with average magnetization pointing between the (+$y$) direction of the applied field and the long axis direction (either (+$z$) or (-$z$)) favored by shape anisotropy, with a modulation along the $x$-direction due to the influence of cubic anisotropy and magnetostatics which promotes a periodic texture. The period of the oscillation in the precursor state is 240 nm, which is in good agreement with our experimentally observed oscillation period. The precursor oscillatory state shown in Fig. 5b is the ground state of the Ni NW magnetization in the presence of a 2.5 kOe magnetic field applied in the $y$-direction.

Two possible magnetic configurations can be obtained from this point. If the field is abruptly removed, the oscillatory precursor configuration is out of equilibrium, and multiple vortices nucleate across the NW with a chirality and period reflecting the magnetization modulation of the precursor state (Fig. 5c). The magnetic configuration thus obtained is consistent with the experimentally obtained PEEM images (Fig. 4) in terms of texture and periodicity ($\lambda$ = 240 nm). Conversely, if the field is gradually reduced from 2.5 kOe toward zero, a single vortex nucleates near an end of the NW at around 2.0 kOe, and expands across the full length of the structure before any other nucleation events take place. As a result, the NW is left in a uniform vortex state with polarity either in the (+$z$) or (–$z$) direction as shown in Fig. 5d, corresponding to the $z$-component of magnetization of the precursor state. The formation of a vortex domain in the Ni NW reduces the magnetostatic energy and aligns spins closer to the <111> axes to reduce the MCA energy.

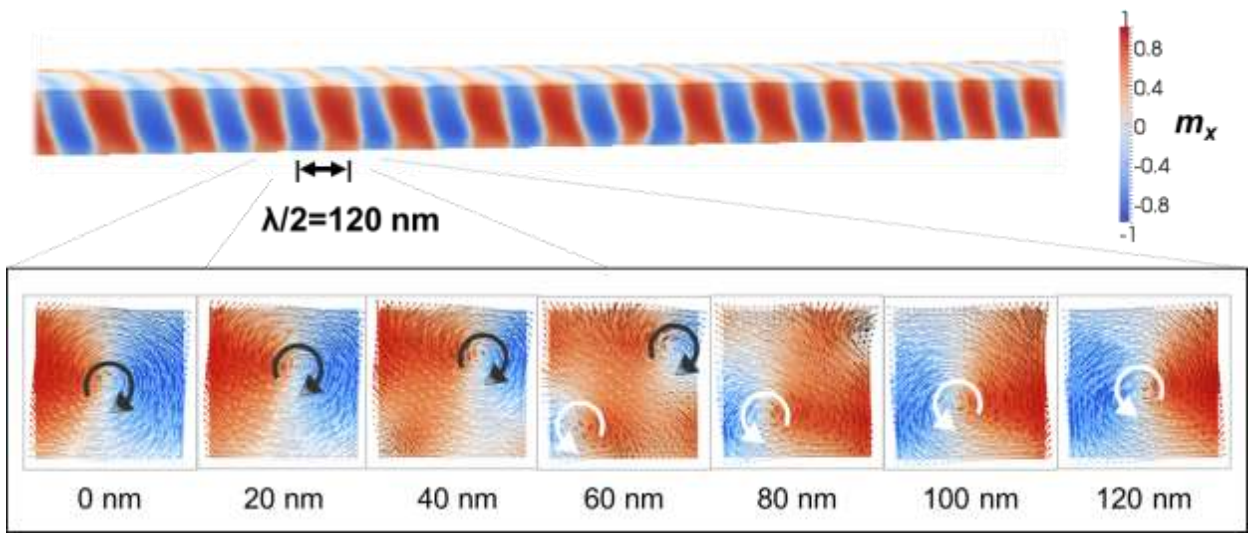

*Figure 6: Exploded slices of the micromagnetic configuration of Ni NW, incrementing in distance along the wire axis showing a half period. Arrows clarify the chirality of vortex domains existing within the NW. Cross sections along the length of the NW in the periodically alternating chirality vortex state. An x-y projection of the spins is shown, but all arrows have a component along the wire axis (towards the reader). The transition between left and right handed configurations involves simultaneous expulsion and nucleation of opposite chirality vortex cores at opposite corners of the NW.*

The magnetic state obtained from the precursor by abruptly removing the field can be described as a periodic array of alternating chirality vortices of same polarity, shown in Fig. 6. The magnetic state is particularly interesting due to the exotic spin configuration which is topologically protected, because once the vortices form, they are effectively locked in, as the collapse of a vortex domain of chirality $\chi$ and merger of two adjacent vortices of chirality $-\chi$ requires the expulsion of a vortex core of chirality $\chi$, and injection of a core of opposite chirality ($-\chi$), which is energetically costly. In this respect, the alternating chirality vortex domain structure is analogous to an array of 360° DWs in a thin strip, where expulsion of a domain necessitates an energetically expensive out-of-plane rotation of magnetic moments[49-52].

Energy calculations show that the alternating chirality vortex state has higher energy than the uniform vortex state, which is the ground state for the modeled cubic NW system at zero field. The presence of multiple domain walls in the former configuration leads to greater overall exchange and anisotropy energy, just as for uniaxial anisotropy systems. The fact that the system relaxes to such a metastable state instead of the ground state when the field is abruptly cut off after reaching the vicinity of 2.5 kOe can be explained by the proximity of the precursor state and the alternating chirality vortex state in configuration space. Namely, when the field is abruptly removed, the out-of-equilibrium magnetization configuration finds and settles in a local energy minimum of the newly formed energy landscape (at zero field) that corresponds to the stable state configurationally most similar to the precursor state.

The stability of the alternating chirality vortex state is a consequence of the energy barrier preventing the collapse of topologically protected alternating chirality vortex domains interior to the NW, and is also due to a particular magnetization configuration at the NW ends, which effectively clogs the NW, preventing vortex domains from escaping the structure at its two extremities. The configuration at the ends is given in Fig. 7, showing a significant portion of the local magnetization oriented in the opposite z-direction to the average magnetization of the NW, i.e., opposite to the polarity direction of the vortices. This configuration is attributed to the interplay of the magnetostatic, cubic anisotropy, and exchange interactions at the discontinuity, which leads to a magnetization configuration at the NW ends that differs from the magnetization pattern interior to the NW. As a result, the vortices cannot escape at the ends without a reconfiguration of magnetization at one or both extremities. The energy barrier associated with such a reconfiguration contributes to the stability of the alternating chirality vortex state by effectively blocking the escape routes of vortices at the two discontinuities.

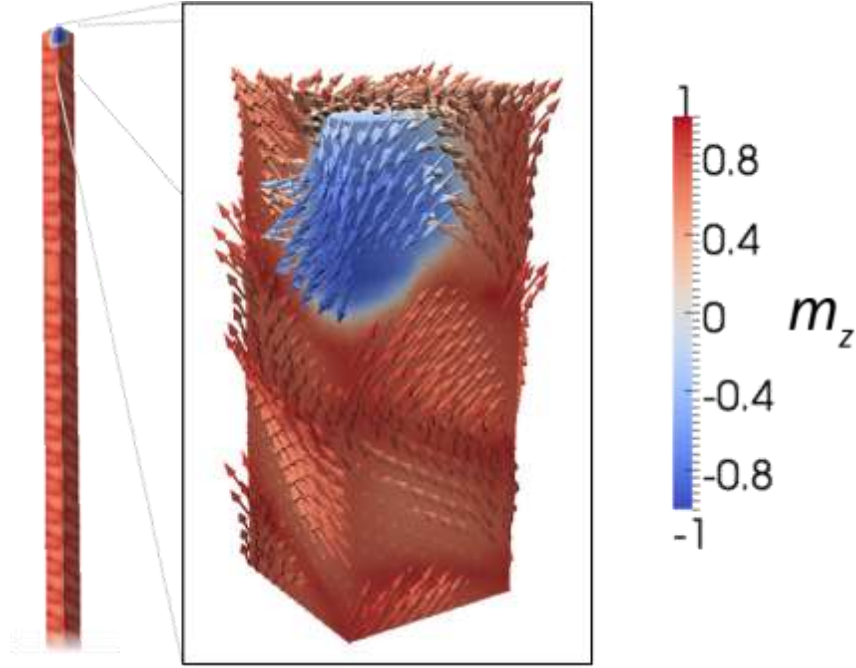

Figure 7: Magnetization configuration at NW end.

An alternative state, intermediate between the uniform vortex ground state and the periodic alternating chirality vortex state, is obtained once the magnetization relaxes to equilibrium following abrupt removal of field at saturation (above 10 kOe) in the *y*-direction, or upon relaxation to equilibrium from a completely random initial state that is obtained by specifying the magnetization vector at each node of the finite element mesh to be in an arbitrary direction at initialization. In both cases, the system is in a highly out-of-equilibrium state as it begins to relax. Consequently, the vortices that first nucleate along the NW (at arbitrary locations) quickly expand on account of the non-equilibrated surrounding magnetization, locally minimizing the energy and ultimately leaving the NW in a stable non-periodic alternating chirality vortex state shown in Fig. 8. The obtained (meta)stable configurations depicted in Fig. 5 c,d and Fig. 8 indicate that the relaxation process and the final equilibrated magnetization configuration of the modeled Ni NW are dependent on field history.

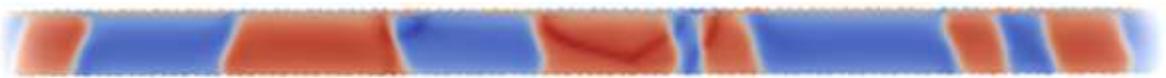

Figure 8: Non-periodic alternating chirality vortex state at *T* = 10 K obtained when the saturating field is abruptly cut of at saturation (~10 kOe).

The anisotropy of single crystal Ni strongly depends on temperature[46]. To study the magnetic response of Ni NW at room temperature, simulations were performed with values of material parameters corresponding to T = 293 K, i.e.: $M_S$ = 450 emu/cm$^3$, $A$=1.0 μerg/cm, $K_1$ = -0.057 Merg/cm$^3$ and $K_2$ = -0.023 Merg/cm$^3$. Due to the much lower magnetic anisotropy energy density at room temperature, the shape anisotropy of the slender NW dominates over cubic anisotropy, and the obtained magnetization at ground state (in the absence of field) is seen to be largely uniform and in the direction of the NW long axis, with moderate nonuniformity at the NW extremities. The absence of periodic texture in the magnetization of the modeled system at $T$ = 293 K is at variance with the oscillations observed experimentally at room temperature using PEEM. The discrepancy suggests a disparity between material properties of the experimentally grown and the modeled Ni NW.

By varying the values of cubic anisotropy energy densities $K_1$ and $K_2$, introducing negative uniaxial anisotropy, or including magnetoelastic effects in the model[53], the interplay between the MCA and shape anisotropy can be tuned to produce a variety of magnetic configurations. Figures 9a,b show a vortex domain formation which develops at room temperature when a negative uniaxial anisotropy is introduced to the system, with anisotropy energy densities $K_u$ = -0.01 Merg/cm$^3$ and $K_u$ = -1.0 Merg/cm$^3$, respectively. Negative uniaxial anisotropy favors magnetization alignment in the plane of the NW cross section, promoting the formation of vortex domains[54].

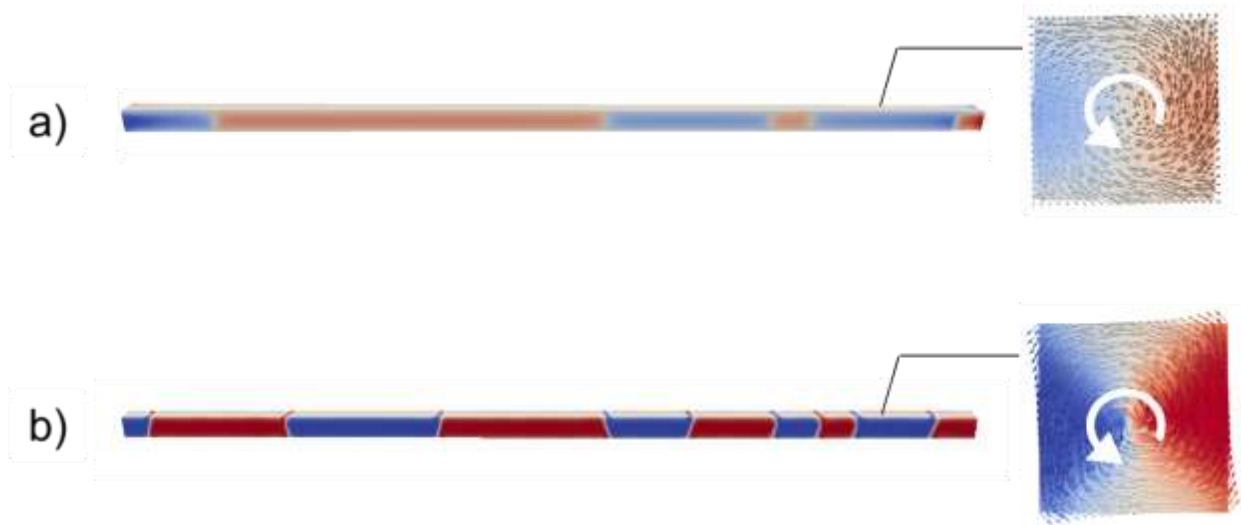

Figure 9: Alternating chirality vortex domains obtained at room temperature, in the presence of negative uniaxial anisotropy, with energy density constants (a) $K_u$ = -0.01 Merg/cm$^3$ and (b) $K_u$ = -1.0 Merg/cm$^3$.

Vortex formation is also seen to result with the introduction of the magnetoelastic interaction[53], assuming uniform compressive biaxial strain in the *xy* basal plane. Such a strain could arise from the synthesis process of the single crystal Ni NWs which grow vertically from the amorphous SiOx coated Si substrate[43, 55]. Figure 10 shows the remanent configuration of the modeled Ni NW assuming magnetostrictive constants $\lambda_{111}$ = -24×10$^{-6}$ and $\lambda_{100}$ = -46×10$^{-6}$, Young's modulus $E$ = 200 GPa, and strain $\varepsilon$ = -0.05. For uniform biaxial compressive strain and magnetostrictive properties corresponding to nickel[56], the <111> directions are the easy magnetization directions, thus favoring vortex formation over uniform magnetization along the NW long axis.

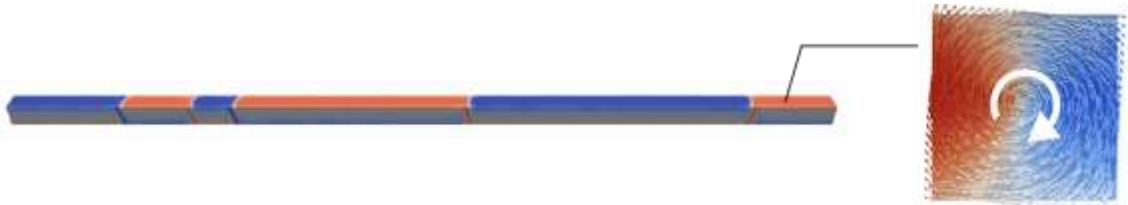

*Figure 10: Alternating chirality vortex domains obtained at room temperature in the presence of uniform biaxial compressive strain ε = 0.05, assuming magnetostrictive constants $\lambda_{111}$ = -24×10$^{-6}$ and $\lambda_{100}$ = -46×10$^{-6}$, Young's modulus E = 200 GPa, typical of nickel.*

Figure 11 shows the spontaneously periodic magnetization pattern which develops in the presence of negative uniaxial anisotropy and in the absence of magnetoelastic effects, when the saturation magnetization is reduced from 450 emu/cm$^3$ to 250 emu/cm$^3$. In this case, vortices are no longer energetically favorable, and the magnetization in the cross sections of the NW is largely uniform (Fig. 11). Due to the accumulation of effective magnetic surfaces charges at the sides of the NW to which the magnetization is normal, a periodic texture develops resulting in flux closure and reduction in magnetostatic energy[57] that is analogous to stripe domains in magnetic thin films. Results shown in Figs. 10-12 indicate that the magnetic response of the studied Ni NW is strongly dependent on material properties of the system.

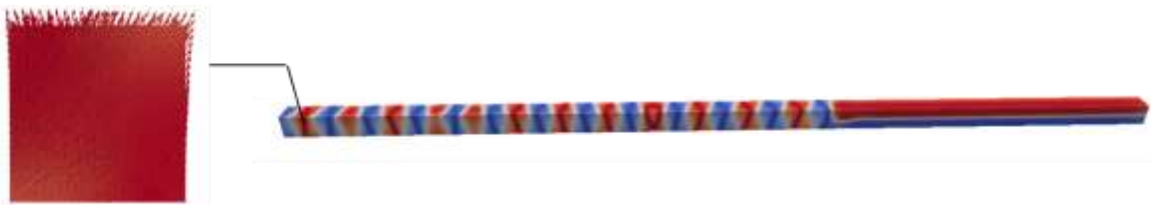

*Figure 11. Periodic array of domains obtained in the presence of negative uniaxial anisotropy ($K_u$ = -1.0 Merg/cm$^3$) at reduced saturation magnetization of $M_s$ = 250 emu/cm$^3$.*

In summary, we have investigated the magnetization configurations arising in single-crystal Ni NWs of dimensions 200 x 200 x 8000 nm. Utilizing magneto-transport and magnetic microscopy techniques, we revealed that the magnetization state of the Ni NW system can exist as a series of alternating domains that are periodically distributed along the nanowire axis, with a period of λ ~ 250 nm. Micromagnetic simulations reveal that an array of alternating chirality topologically protected vortex domains with a period closely matching experimental observation (λ ~ 240 nm) is a possible remanent state. Vortex formation in single crystal Ni NWs is attributed to cubic anisotropy of the material system, which favors magnetization to point in the easy <111> directions, and its reduced structural dimensions. Simulations show that other remanent states are also possible depending on the field history. The ground state is found to correspond to a single uniform vortex domain extending along the length of the NW. Because of a much smaller cubic anisotropy of Ni at room temperature, shape anisotropy dominates MCA, resulting in largely uniform NW magnetization oriented longitudinally along the NW. With introduction of negative uniaxial anisotropy or in the presence of magnetoelastic effects under the assumption of uniform compressive biaxial strain in the basal plane, vortices are seen to reappear at $T$ = 293 K. Rapid quenching of the NW temperature during the synthesis process and strain buildup could bias NW magnetization toward the periodic alternating chirality vortex metastable state observed in simulations. The alternating chirality topologically protected vortex domains could be found interesting for employment in magnetic memory and logic technologies where bits of information are represented by magnetic domains that can be stored in structures of reduced dimensions and manipulated and transmitted by applied fields, strains, and/or spin polarized currents.

This work was supported by the National Science Foundation through grant award number: DMR-0906957. This research used resources of the Advanced Light Source, which is a DOE Office of Science User Facility under contract number: DE-AC02-05CH11231.

* Current affiliation: Anzu Partners, LLC, La Jolla, California, 92037, USA

† Current affiliation: Global Forecasting, Gartner, Inc., Stamford, Connecticut 06902, USA